# The Green Advantage: Analyzing the Effects of Eco-Friendly Marketing on Consumer Loyalty

**Erfan Mohammadi**

Faculty of Entrepreneurship, University of Tehran, Tehran, Iran

**MohammadMahdi Barzegar**

Sheldon B. Lubar College of Business, University of Wisconsin Milwaukee

barzega3@uwm.edu

**Mahdi Nohekhan**

Executive Management Group, Islamic Azad University, Shahrood Branch

**Abstract**

The idea that marketing, in addition to profitability and sales, should also consider the consumer's health is not and has not been a far-fetched concept. It can be stated that there is no longer a way back to producing environmentally harmful products, and gradually, governmental pressures, competition, and changing customer attitudes are obliging companies to adopt and implement a green marketing approach. Over time, concepts such as green marketing have penetrated marketing literature, making environmental considerations one of the most important activities of companies. For this purpose, this research examines the effects of green marketing strategy on brand loyalty (case study: food exporting companies). The population of this study consists of 345 employees and managers of companies like Kalleh, Solico, Pemina, Sorben, Mac, Pol, and Casel, out of which 182 were randomly selected as a sample using Cochran's formula. This research is practical; the required data were collected through a survey and questionnaire. The research results indicate that (1) green marketing strategy significantly affects brand loyalty. (2) Green products have a significant positive effect on brand loyalty. (3) Green promotion has a significant positive effect on brand loyalty. (4) Green distribution has a significant positive effect on brand loyalty. (5) Green pricing has a significant positive effect on brand loyalty

**Introduction**

Over the past few decades, competition among manufacturers has continuously focused on the quality and characteristics of products, increasing the risk of product obsolescence. However, focusing on quality is only feasible in business through a trade-off between quality and cost. Traditionally, higher quality increases production costs for various reasons, such as using high-quality materials and different operational expenses. (Wood Herkom et al., 2015). Two factors influence consumer purchasing decisions for green products. One set of these factors relates to internal aspects of consumers, such as realizing environmental responsibilities, striving to gain knowledge, and the desire to conserve resources and reduce environmental impacts. The other group of factors pertains to external aspects of consumers. These include the social aspect, product characteristics (product quality, safety, price, promotion, and their effects on human health), and consumer behavior, which in reality is the actual behavior resulting from consumers' regular habits, their awareness of the product, and situational factors such as the company's advertising methods (Khar, 2015).

In sustainability literature, green marketing refers to the practices, policies, and marketing programs that specifically address environmental concerns and aim to generate revenue and provide outcomes that fulfill individual and organizational goals for a product or production line (Carsten & Iris, 2014). Therefore, green marketing programs are designed to minimize negative impacts on the natural environment. On the other hand, appropriate positioning begins when a company offers something genuinely different from its competitors, thereby creating value for customers (Kotler, 1999).

Increasing environmental pollution: Given the increasing pollution of the environment, companies must take steps towards the green movement and green marketing to prevent this. Environmental pollution has caused social pressures from consumers and government policies to steer companies toward going green (Kateb & Helsen, 2004). Among the companies that have changed their policies due to social pressures is Exxon (the largest oil pipeline company). It has changed its policies under social pressure and established an environmental protection institute and two local parks in Cameroon—several internal factors pressure companies to implement green activities.

The first and most important reason is the cost factor. This means that being green can lead to greater efficiency, benefits, and financial savings. This implies using fewer inputs and reducing waste and pollution (Plonsky & Rosenberg, 2001).

The second internal factor is the company's philosophy. When companies value environmental goals as much as other company goals and incorporate environmental issues into their company philosophy, the discussion of being green intertwines with the company's strategies and then blends it with its tactical activities.

The third internal factor is creating a competitive position in the market. Companies that consider environmental issues in their marketing and product production processes create a competitive position compared to their rivals. Therefore, it can be said that adhering to green marketing principles gives consumers a better view of the company (Kotler, 1999). Ken Peattie states that environmental issues have led to the creation of a new approach, which he calls ecological positioning (Peattie, 2001). According to this approach, companies should emphasize environmental issues for positioning their products.

Given the increasing customer awareness of environmental issues and the strict laws introduced by national governments, especially in developed industrial countries, companies and manufacturers must inevitably comply with environmental standards when producing goods or providing services. Western studies show that the environmental awareness of customers in America and Western Europe has been growing over the past decade, and recently, green consumption has also started in Asian regions. A key opportunity arising from using green marketing is a new playing field with few competitors. By going green, a company can gain several advantages, including improved resource efficiency, reduced company structural costs, and improved competitive positioning. Moving towards green has the following benefits:

- It leads to improved resource efficiency, thereby reducing costs in the company structure and enhancing the company's competitive position relative to competitors.
- Being green enables a company to stand out from competitors by offering new products in new markets or providing additional benefits for current products. This increases the company's value to customers, enhances customer loyalty, and ultimately increases profitability (Peattie, 2001).

Customer loyalty creates a competitive advantage for the company because loyal customers save marketing costs. This is because, as experts say, the cost of attracting new customers is about six times that of retaining current ones (Karbasi et al., 2011). Various internal and external factors influence customer loyalty to an organization, and their impact varies from one organization to another. Precise recognition of these factors and determining their impact is of great importance in assisting managers in making correct decisions. The literature on service marketing focuses on key concepts and constructs such as justice, customer orientation, service quality, satisfaction, and brand loyalty (Safarzadeh & Esmaili, 2011).

Creating brand loyalty requires investment in marketing programs, especially for potential and current customers. These marketing activities have influenced customers' mindsets and led to various outcomes. All different areas of customer contact with the brand provide an opportunity to create a favorable attitude and increase brand loyalty, with the main area of customer interaction being the sales encounter. Since a salesperson is the only person who interacts with the customer, they can play a key role in the customer's experience and brand evaluation (Karbasi et al., 2011).

Loyalty is a positive attitude towards a product that arises from repeated use, which psychological processes can explain. In other words, repeat purchasing is not just a voluntary reaction but the result of psychological, emotional, and normative factors (Safarzadeh & Esmaili, 2011).

The concept of loyalty was first introduced in the 1940s. At that time, loyalty was a one-dimensional concept. Later, in 1944 and 1945, two separate concepts about loyalty emerged: 'brand preference,' later known as attitudinal loyalty, and 'market share,' later presented as behavioral loyalty. Thirty years later, the concept of loyalty entered academic literature, and researchers found that loyalty could be a combination of attitudinal and behavioral loyalty (Safarzadeh & Esmaili, 2011).

Today, brands and trademarks are among the most important marketing topics facing all companies, including commercial companies. This importance has reached such a level that brand management has become an independent field in academic circles. Brands are considered economically and strategically as one of companies' most valuable assets, even arguably their most valuable assets. In recent years, researchers have concluded that the real value of products and services for companies is not within the products and services themselves but in the minds of actual and potential customers, and it is the brand that creates real value in the minds of customers (Keller, 2014).

One of the most important goals of brand owners for brand development is to create customer loyalty towards their brand. Loyal customers are considered a reliable source of profitability for companies, providing them with a suitable safety margin in an uncertain environment. One of the ways to create customer loyalty to a brand is to develop an appropriate brand identity. Brands distinguish themselves from competitors by their identity (Keller, 2014).

Brand loyalty means repeat purchasing, which is due to psychological processes. Keller states that brand loyalty in the past was often measured by repeat purchasing behaviors, while customer loyalty can be more broadly considered than just simple purchasing behaviors. Brand loyalty consists of two components: behavioral loyalty and attitudinal loyalty. Customer loyalty to a brand leads to positive word-of-mouth advertising, creating significant barriers to entry for competitors, empowering the company to respond to competitive threats, generating more sales and revenue, and reducing customer sensitivity to competitors' marketing efforts. Many loyal customers to a brand are considered an asset of the company and are recognized as the main indicator of brand equity. Loyal customers are less sensitive to price changes compared to non-loyal customers. Loyalty leads to repeat purchasing of consumer goods. In marketing literature, brand loyalty is often synonymous with concepts such as 'repurchase,' 'preference,' 'commitment' (Bai & Vahedian, 2023), and 'adherence,' and these terms are used interchangeably (Hessari, 2023; Sahin et al., 2011).

Brand loyalty can be defined as the extent to which a customer has a positive attitude towards a brand, their commitment to the brand, and their intention to continue purchasing. Brand loyalty creates a commitment to repurchase. According to Tellis, the impact of brand loyalty is greater than other variables. Atilgan confirmed this finding and stated that loyalty is the only factor directly affecting brand equity. In Aaker's brand equity model, loyalty is mentioned as a commercial lever to reduce marketing costs, attract new customers (in terms of creating awareness and reassurance), and time to respond to competitive threats. He also defined the loyalty index to a brand in two ways: (1) the amount a customer pays for a brand compared to other brands with similar benefits. (2) Direct measurement of satisfaction and intent to repurchase or order a product or service again. Oliver (1997) described brand loyalty as a deep commitment to repurchase or use a product or service in the future despite environmental effects and marketing efforts that could change behavior. Papo (2006) stated that a customer's first choice of a brand is a reason for their loyalty and considers two dimensions of loyalty: attitudinal and behavioral loyalty. The consumer's purchasing experience expresses the behavioral loyalty dimension and forms the reason for a specific purchasing habit, measuring the actual purchasing behavior of a consumer towards a product. From an attitudinal perspective, a positive commitment is created between the consumer and the brand, and this attitude creates trust between the brand's reputation and the consumer's benefits, expressing the consumer's attitude towards a product.

Gil also stated that loyalty directly increases brand equity, and other variables also affect brand equity both directly and indirectly through the loyalty variable. Yu and others' research (2000) confirmed this point. However, in 2000, a study focusing on cultural differences between the United States and Korea across 12 different brands showed that brand loyalty positively affected brand equity. However, this effect differed in both markets (Sahin et al., 2011).

According to Lee and Beck (2008), brand equity is nothing but loyalty and the mental image customers have of the brand. Customer loyalty is more than just repeating purchase occurrences. Dacko (2008) defines brand loyalty as the degree of preference for a brand by a customer over close substitutes. Aaker (1991) defines brand loyalty as a customer's attachment and belonging to a brand. In other words, brand loyalty is the likelihood of not turning away from one brand to another, especially when that brand changes goods, prices, or other factors. Brand equity is primarily derived from brand loyalty, and this loyalty, in turn, stems from customers' experiences with the brand. Oliver (1999) presented a four-stage model for loyalty, in which customers first become cognitively loyal to a brand, then develop emotional loyalty, followed by attitudinal loyalty, and finally demonstrate behavioral loyalty to the brand. Also, brand loyalty is examined from attitudinal (having a positive attitude towards the product and brand) and behavioral (showing purchasing behavior). From a behavioral perspective, brand loyalty is defined as repeat purchasing and brand preference. From an attitudinal perspective, it is preferring and selecting it as the first purchasing choice, having a positive feeling towards continuing the relationship with the brand, and sharing this enjoyable experience with others (Sahin et al., 2011).

The more customers trust and feel assured about products under a specific brand, due to the nature of individuals' risk aversion and reducing their purchasing risk, the more they try to buy products of the same brand and show less inclination towards products of other brands, which means customer loyalty to the brand. Trust can be defined as the customer's assurance that leads them to rely on the seller until the promised services are delivered (Sahin et al., 2011).

**Dimensions of Brand Loyalty**

To broadly illustrate the concept of customer loyalty to a brand, it includes various dimensions, such as:

**Cognitive Loyalty**: This leads to customer behavior and encompasses all aspects where the customer expects satisfactory performance from the product. Based on available information, it determines the superiority of one option over others for the customer. Loyalty at this stage is based on the customer's beliefs and cognition through prior knowledge or information from recent experiences, essentially relating to rational decision-making.

**Emotional Loyalty**: Involves affection and attitude towards a brand. This affection and attitude, resulting from multiple satisfactions derived from using the brand, turn into positive feelings towards that brand. It refers to the commitment and trust of the customer. It encompasses all aspects where the consumer, based on emotional evaluation, chooses a brand that may not always align with rational criteria (Philo & Funk, 2008).

Loyal consumers tend to pay more for branded products because they feel they create more value for them than others. The intention to repurchase is also an indicator of brand loyalty. The literature on measuring loyalty shows an evolutionary development that started with concepts based on behavior but now includes approaches based on attitude, cognition, and values. Behavioral approaches operationalize loyalty in 4 ways: First, by measuring the actual consumption of goods and services, usually combining volume and frequency of purchases over specific periods. Ehrenberg (1998) observed that patterns from such measurements greatly assist marketers in identifying heavy users and repeat buyers. Second, consider criteria or standards within a defined market or retail location. Third, criteria or standards are based on the likelihood of repeat purchases. Fourth, criteria or standards in fact examine when customers switch to other brands. In addition to these, other researchers have considered various criteria such as brand familiarity, ease and consumption experience, social value, perception and self-concept, perceived value, and satisfaction to measure brand loyalty (Bai et al., 2023; Safarzadeh & Esmaili, 2011).

Rostamzadeh and Mohammadi Saiban (2016) investigated the impact of environmental effects on green marketing performance; case study: Shirin Asal Company. Over the past decades, concern about the environment has become a significant public issue and a critical topic in academic research. Given the rising concern, the market for ecological products is growing globally. Consequently, green marketing activities are increasing in many parts of the world, offering products to green consumers who make their purchasing decisions based partly on personal environmental criteria. These activities significantly impact increasing consumer knowledge and changing consumer attitudes toward purchasing green products. Thus, considering that the food industry reflects the economic and industrial development of a country and its extensive operations from production to distribution and consumption play a significant role in driving the economy of any country, this research aims to examine the impact of environmental strategy on green marketing performance in the food industry of Shirin Asal. Structural equation modeling was used for causal assessment and to check the reliability and validity of the model. The presented model and its results show that environmental strategy affects financial performance, market performance, and service quality.

Imam Gholi (2014) examined the role of customer relationship management in marketing green products. This research aimed to investigate the role of customer relationship management in green marketing and cosmetic and hygiene green products. The statistical population of this research was the cosmetics and hygiene sector across the city of Rasht, from which a sample of 131 sellers of cosmetics and hygiene products was selected. Sellers' attention to selling green products and customers' attention to buying green products was examined. After explaining the theoretical and research bases related to customer relationship management and green marketing, the role of customer relationship management on the four green marketing mix elements (product, price, distribution, promotion) was examined. After preparing a researcher-made and standardized questionnaire controlling its validity and reliability using Cronbach's alpha formula and conducting a normality test, data were collected using SPSS15 software, and the hypotheses of this research were analyzed using independent two-sample t-tests and variance analysis. In the end, the elements of the green marketing mix were ranked using the Friedman test, and it was determined that there is a significant relationship between customer relationship management and the elements of the green marketing mix.

Osama (2011) investigated green marketing: marketing strategies of Swedish energy companies. Commercial companies have become aware of the need to preserve the environment and have included marketing strategies for environmental protection as a social responsibility in their agenda. This article analyzes the reasons why companies have adopted green marketing (related to producing products that are not harmful to the environment) the research methods used, and the results obtained. This article is primarily based on related writings and a review of empirical studies to understand the importance of incorporating environmentally-based marketing into companies. It also states that environmental labeling on products distinguishes environmentally friendly producers from ordinary producers. The key findings of this article are that large companies cannot be completely detached from green business and must equally participate in social programs. These companies have been given decision-making power based on how much they are willing to incorporate green marketing strategies into their overall company plans. This environmental model is derived from a study of existing business literature.

**Research Design**

**Research Methodology and Objective**: This study is applied in terms of its objective and descriptive survey in terms of its research method, categorized as correlational.

**Sampling Method and Data Collection**: The population of this study includes managers and employees of food export companies such as Kalleh, Solico, Pemina, Sorben, Mac, Pol, and Casel, who were active in the years 2015-2016. According to the obtained statistics, their total number is 345, of which approximately 182 individuals were selected based on Cochran's formula.

**Statistical Methods and Data Analysis Approach**: In this research, the Green Marketing Strategy questionnaire, derived from Imam Gholi's research (2014) (which includes 7 questions on green products, 8 on green promotion, 3 on green distribution, 5 on green pricing), and the Aaker (1991) Brand Loyalty questionnaire (which includes seven questions) were used. Since this research involves a sample of the population and the data are based on a 5-point Likert scale ranging from "Strongly Disagree" to "Strongly Agree," the scoring of the questions was calculated from 1 to 5. After collection, the data from the questionnaires were transferred to the raw data sheets of SPSS 19 software and analyzed using descriptive statistics (frequency, percentage, tables) and inferential statistics (Pearson correlation).

**Reliability and Validity of the Research Questionnaires**: To assess the validity of the measurement tools used in this research, the opinions of university professors and experts were utilized. Also, since the items considered in this questionnaire are based on standard questionnaires, the questionnaire is deemed valid. The current study conducted a preliminary sample including 30 pre-test questionnaires. Then the reliability coefficient was calculated using the data obtained from these questionnaires and SPSS software, using Cronbach's alpha method. The Cronbach's alpha coefficient for the Green Marketing Strategy questionnaire is 0.972 for the components of Green Marketing Strategy (Green Product 0.945, Green Promotion 0.881, Green Distribution 0.806, Green Pricing 0.912), and for the Brand Loyalty questionnaire (0.929). Therefore, it can be said that the questionnaires have high reliability, internal consistency, and validity.

**Conceptual Model of Research and Hypotheses**: In this research, following the model of Hick and Yildan (2013) and Moskovi et al. (2015), hypotheses are tested. The conceptual model of the present research is researcher-made and results from reviewing various models and literature (Figure 1).

Figure 1: Conceptual Model of Research

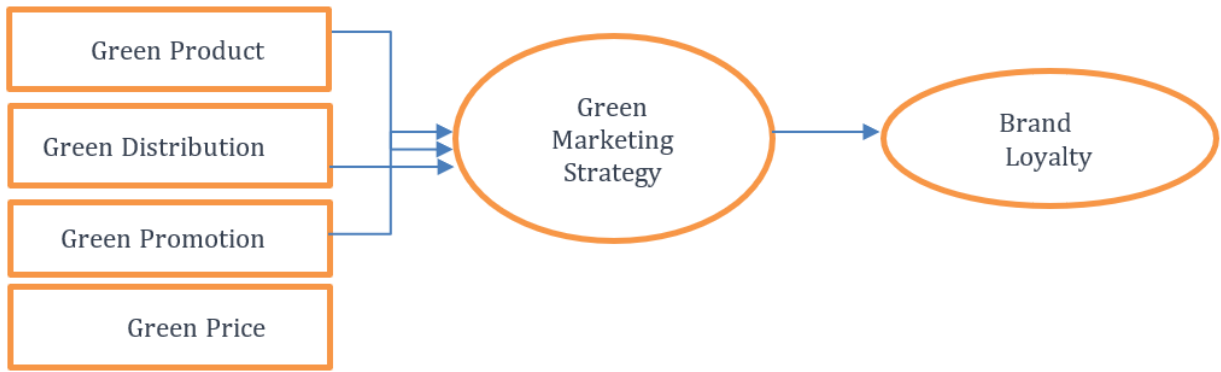

**Main Hypothesis of the Research**: Green marketing strategy significantly affects brand loyalty.

**Sub-Hypothesis 1**: Green products have a significant effect on brand loyalty.

**Sub-Hypothesis 2**: Green promotion has a significant effect on brand loyalty.

**Sub-Hypothesis 3**: Green distribution has a significant effect on brand loyalty.

**Sub-Hypothesis 4**: Green pricing has a significant effect on brand loyalty.

**Findings**

The frequency and percentage distribution of the research sample group based on gender, educational qualification, and Work Experience are presented in Table (1).

Table 1: Frequency and percentage distribution of the research sample group

| Variable | Subgroups | Frequency | Percentage Frequency |
|---|---|---|---|
| Gender | Male | 103 | 56.60% |
|  | Female | 79 | 43.40% |

| Age | Less than 30 | 53 | 29.10% |
|---|---|---|---|
| | 31 to 40 | 61 | 33.50% |
| | 41 to 50 | 46 | 25.30% |
| | More than 50 | 22 | 12.10% |
| Education | Diploma and below | 31 | 17.00% |
| | Associate degree | 43 | 23.60% |
| | Bachelor's degree | 59 | 32.40% |
| | Graduate degree | 49 | 26.90% |
| Work Experience | Less than 1 year | 21 | 11.50% |
| | 1 to 5 years | 34 | 18.70% |
| | 6 to 10 years | 78 | 42.90% |
| | More than 10 years | 49 | 26.90% |

As indicated in Table 1, the majority of the study sample in terms of gender are men; in terms of educational level are those with a bachelor's degree; in terms of work experience are those with 6-10 years, and in terms of age are those between 31-40 years old.

Table 2: Pearson Correlation Test (Main Hypothesis)

| | | Green Marketing Strategy | Brand Loyalty |
|---|---|---|---|
| Green Marketing Strategy | Pearson Correlation Coefficient | 1 | .890 |
| | Significance Level | | .000 |
| | Sample Size | 182 | 182 |
| Brand Loyalty | Pearson Correlation Coefficient | .890 | 1 |
| | Significance Level | .000 | |
| | Sample Size | 182 | 182 |

The results of the Pearson correlation test shown in Table 2 indicate that the significance level is zero and less than 0.05, demonstrating a significant relationship between green marketing strategy and brand loyalty. Therefore, the main hypothesis of the research is confirmed. Also, the Pearson correlation coefficient is (0.890), indicating a significant positive effect of green marketing strategy on brand loyalty.

Table 3: Pearson Correlation Test (First Sub-Hypothesis)

| | | Brand Loyalty | Green Product |
|---|---|---|---|

|  |  | Brand Loyalty | Green Product |
|---|---|---|---|
| **Brand Loyalty** | Pearson Correlation Coefficient | 1 | .738 |
|  | Significance Level |  | .000 |
|  | Sample Size | 182 | 182 |
| **Green Product** | Pearson Correlation Coefficient | .738 | 1 |
|  | Significance Level | .000 |  |
|  | Sample Size | 182 | 182 |

The results of the Pearson correlation test presented in Table 3 indicate that the significance level is zero and less than 0.05, demonstrating a significant relationship between green products and brand loyalty. Therefore, the first subsidiary hypothesis of the research is confirmed. Additionally, the Pearson correlation coefficient is (0.738), indicating a significant positive effect of green products on brand loyalty.

Table 4: Pearson Correlation Test (Second Sub-Hypothesis)

|  |  | Brand Loyalty | Green Promotion |
|---|---|---|---|
| **Brand Loyalty** | Pearson Correlation Coefficient | 1 | .898 |
|  | Significance Level |  | .000 |
|  | Sample Size | 182 | 182 |
| **Green Promotion** | Pearson Correlation Coefficient | .898 | 1 |
|  | Significance Level | .000 |  |
|  | Sample Size | 182 | 182 |

The results of the Pearson correlation test in Table 4 show that the significance level is zero and less than 0.05, indicating a significant relationship between green promotion and brand loyalty. Therefore, the second subsidiary hypothesis of the research is confirmed. Additionally, the Pearson correlation coefficient is (0.898), indicating a significant positive effect of green promotion on brand loyalty.

Table 5: Pearson Correlation Test (Third Sub-Hypothesis)

|  |  | Brand Loyalty | Green Distribution |
|---|---|---|---|
| **Brand Loyalty** | Pearson Correlation Coefficient | 1 | .930 |
|  | Significance Level |  | .000 |
|  | Sample Size | 182 | 182 |
| **Green Distribution** | Pearson Correlation Coefficient | .930 | 1 |
|  | Significance Level | .000 |  |
|  | Sample Size | 182 | 182 |

The results of the Pearson correlation test in Table 5 show that the significance level is zero and less than 0.05, indicating a significant relationship between green distribution and brand loyalty. Therefore, the third subsidiary hypothesis of the research is confirmed. Additionally, the Pearson correlation coefficient is (0.930), indicating a significant positive effect of green distribution on brand loyalty.

Table 6: Pearson Correlation Test (Fourth Sub-Hypothesis)

|  |  | Brand Loyalty | Green Pricing |
|---|---|---|---|
| **Brand Loyalty** | Pearson Correlation Coefficient | 1 | .768 |
|  | Significance Level |  | .000 |

| | Sample Size | 182 | 182 |
|---|---|---|---|
| **Green Pricing** | Pearson Correlation Coefficient | .768 | 1 |
| | Significance Level | .000 | |
| | Sample Size | 182 | 182 |

The results of the Pearson correlation test in Table 6 indicate that the significance level is zero and less than 0.05, demonstrating a significant relationship between green pricing and brand loyalty. Therefore, the fourth subsidiary hypothesis of the research is confirmed. Additionally, the Pearson correlation coefficient is (0.768), indicating a significant positive effect of green pricing on brand loyalty.

Table 7: Research Hypotheses Examination Results

| Hypothesis | Pearson Coefficient | Confirmation/Rejection | Type of Relationship |
|---|---|---|---|
| Main | 0.890 | Confirmed | Positive |
| Sub 1 | 0.738 | Confirmed | Positive |
| Sub 2 | 0.898 | Confirmed | Positive |
| Sub 3 | 0.930 | Confirmed | Positive |
| Sub 4 | 0.768 | Confirmed | Positive |

**Discussion and Conclusion**

The present research is an effort to analyze the impact of green marketing strategy on brand loyalty in Tehran's food export companies. The results of the hypothesis testing showed that green marketing strategy, products, promotion, distribution, and pricing have a significant positive effect on brand loyalty. These results are in line with the research of Hick and Yildan (2013), Moghaddam and Amirhosseini (2015), Hessari et al. (2023) and Souri et al. (2015).

It is recommended that in the company's marketing programs, various methods of introducing green products, such as providing samples as freebies and sampling, be used so that customers, while physically engaging (through the five senses), become familiar with the features of the products and their distinction from other products.

It is also suggested that the methods the company uses to protect and support the environment in its activities be presented using tangible models and mock-ups, or through photographs, brochures, and advertising posters. These should clearly explain the benefits of using environmentally friendly products, making them easily understandable and explicable to customers of different cultural levels and consumers.

Creating, participating in, and supporting environmental organizations and active NGOs in this field can significantly contribute to brand awareness by linking industry and the environment and moving towards promoting green industry.

It is recommended to use technical knowledge to package products in special and biodegradable containers and offer them to customers. This not only increases the shelf life of the products and ensures high-quality products reach the customers, but also the packaging, while not polluting the environment or quickly decomposing and returning to nature, assures customers of the brand's commitment to environmental preservation.